\documentclass[journal=jpccck,manuscript=article]{achemso} % ,layout=twocolumn produces similar looking pdf as in journal, http://www.latex-community.org/forum/viewtopic.php?f=47&t=20538
\usepackage{amsmath}
\usepackage{amssymb}
\usepackage{color}
\usepackage{booktabs}
\usepackage[colorlinks=true,linkcolor=blue]{hyperref}
\usepackage{longtable}

\newcommand{\comment}[1]{} 

\author{Peter Cendula}
\email{cend@zhaw.ch}
\affiliation[Zurich University of Applied Sciences]
{Institute of Computational Physics, Zurich University of Applied Sciences (ZHAW), Wildbachstrasse~21, 8401~Winterthur, Switzerland}

\author{S. David Tilley}
\affiliation[Ecole Polytechnique F\'ed\'erale de Lausanne]
{Laboratory of Photonics and Interfaces, Ecole Polytechnique F\'ed\'erale de Lausanne, EPFL-SB-ISIC-LPI, Station~6, 1015~Lausanne, Switzerland}

\author{Sixto Gimenez}
\affiliation[University Jaume I]
{Photovoltaics and Optoelectronic Devices Group, Departament of Physics, University Jaume I, 12071~Castellon, Spain}

\author{Juan Bisquert}
\affiliation[University Jaume I]
{Photovoltaics and Optoelectronic Devices Group, Departament of Physics, University Jaume I, 12071~Castellon, Spain}

\author{Matthias Schmid}
\affiliation[Zurich University of Applied Sciences]
{Institute of Computational Physics, Zurich University of Applied Sciences (ZHAW), Wildbachstrasse~21, 8401~Winterthur, Switzerland}

\author{Michael Gr\"atzel}
\affiliation[Ecole Polytechnique F\'ed\'erale de Lausanne]
{Laboratory of Photonics and Interfaces, Ecole Polytechnique F\'ed\'erale de Lausanne, EPFL-SB-ISIC-LPI, Station~6, 1015~Lausanne, Switzerland}

\author{J\"urgen O. Schumacher}
\affiliation[Zurich University of Applied Sciences]
{Institute of Computational Physics, Zurich University of Applied Sciences (ZHAW), Wildbachstrasse~21, 8401~Winterthur, Switzerland}

\title[\texttt{achemso} demonstration]
{Calculation of Energy Band Diagram of a Photoelectrochemical Water Splitting Cell}
\begin{document}
%%%%%%%%%%%%%%%%%%%%%%%%%%%%%%%%%%%%%%%%%%%%%%%%%%%%%%%%%%%%%%%%%%%%%
%% The manuscript does not need to include \maketitle, which is
%% executed automatically.  The document should begin with an
%% abstract, if appropriate.  If one is given and should not be, the
%% contents will be gobbled.
%%%%%%%%%%%%%%%%%%%%%%%%%%%%%%%%%%%%%%%%%%%%%%%%%%%%%%%%%%%%%%%%%%%%%
\begin{abstract}

A physical model is presented for a semiconductor electrode of a photoelectrochemical (PEC) cell, accounting for the potential drop in the Helmholtz
layer. Hence both band edge pinning and unpinning are naturally included in our description. The model is based on the continuity equations for charge carriers and direct charge transfer from the energy bands to the electrolyte. A quantitative calculation of the position of the energy bands and the variation of the quasi-Fermi levels in the semiconductor with respect to the water reduction and oxidation potentials is presented. Calculated current-voltage curves are compared with established analytical models and measurement. Our model calculations are suitable to enhance understanding and improve properties of semiconductors for photoelectrochemical water splitting. 
  
%   Rate constants for water oxidation are extracted from experimental data, which is achieved by comparing measured and simulated polarization curves of the one-electron redox couple [Fe(CN)$_6$]$^{3-/4-}$ for various light intensities.
%   (low photocurrent 0.2-0.4mA/cm$^2$).  H$_2$O$_2$ (much higher photocurrent order mA/cm$^2$)
  
  %The research community can use the software free of charge. 
\end{abstract}


% \ToDo{
% Proposed journal submission : Journal of Physical Chemistry C
% %Referees:  H.J. Lewerenz (Caltech, Berlin), B.Dam (Delft), R. van de Kroel (Berlin).
% }

%OPTION 1 (Preferred)
% Submit your own Manuscript PDF File—and provide your native TeX/LaTeX manuscript package as a ZIP archive.Include the main body of the TeX/LaTeX document (i.e., a file ending with .tex or .ltx), all files referenced by the main TeX/LaTeX document (including other .tex files), and any graphics files, compressed into a single .zip file.
% File Designation 	Files for Submission
% Manuscript File 	Submit a ZIP Archive of all TeX files, graphics, and referenced files.
% Manuscript PDF File 	Submit your own author-generated PDF.

\paragraph{Introduction}

Research on hydrogen production by photoelectrochemical (PEC) cells is propelled by the worldwide quest for capturing, storing and using solar energy instead of decreasing fossil energy reserves. Hydrogen is widely considered as a key solar fuel of the future \cite{lewis_powering_2006}. Hydrogen is also part of power to gas conversion systems developed to resolve intermittency in the wind and solar energy production \cite{schiermeier_renewable_2013}. Although a PEC/photovoltaic cell with 12.4\% efficiency was demonstrated with GaInP$_2$/GaAs \cite{khaselev_monolithic_1998}, decreasing its cost and increasing its lifetime are still under way. An alternative approach often pursued is to use abundant and cheap metal oxides as a viable class of semiconductor materials for PEC electrodes \cite{van_de_krol_n-si/n-fe2o3_2013,sivula_solar--chemical_2013,abdi_efficient_2013-1}. However, their recombination losses, charge carrier conduction and water oxidation properties 
need to be understood and optimized both by measurement and numerical simulation \cite{krol_photoelectrochemical_2011}.

Several approaches for a mathematical analysis of semiconductor electrodes can be found in the literature, including analytical \cite{gartner_depletion-layer_1959,wilson_model_1977} and numerical models \cite{reichman_currentvoltage_1980,andrade_transient_2011} of PEC cells. An extensive numerical study of PEC behavior of Si and GaP nanowires was recently conducted with commercial software \cite{foley_analysis_2012}. Since surface states play a major role for many semiconductors, corresponding models were also developed to analyze their effect on electrochemical measurements \cite{peter_surface_1984, klahr_water_2012,bertoluzzi_equivalent_2012}. On the PEC system level, models of the coupled charge and species conservation, fluid flow and electrochemical reactions were recently developed \cite{carver_modelling_2012,haussener_simulations_2013}. The latter studies revealed how PEC systems should be designed with minimal resistive losses and low crossover of hydrogen and oxygen by use of a non-permeable 
separator. 

Almost every publication on PEC cells features a schematic energy band diagram of a PEC cell, mostly sketched by hand from basic physical understanding described in textbooks on electrochemistry \cite{salvador_semiconductors_2001,memming_semiconductor_2008,krol_photoelectrochemical_2011}. Although such sketches might be qualitatively correct, numerical calculations of the charge carrier transport might reveal additional features not captured by the sketches. We are aware that the development of numerical calculations is frequently hindered by the complicated physical processes in the actual materials and lack of measurements of parameter values for these processes \cite{peter_energetics_2013}. In spite of these obstacles, we think that the recent advent of user-friendly numerical software and advanced measurement techniques could fill the gap between experimental and numerical approaches if experimentally validated models are developed.

% Models of impedance spectroscopy based on kinetic model \cite{upul_wijayantha_kinetics_2011} or equivalent circuits \cite{klahr_water_2012,bertoluzzi_equivalent_2012}

%also park bard>2013 continuity equations with SRH recombination for BiVO4. Experimental results shown in Figure 2 for sulfite oxidation of BiVO4 were used to fit the simulation parameters, i.e., the hole and electron transfer rate constants (kf and kb in equations (14) and (15)), the surface recombination rate constants (kSurRec), the electron and hole mobility (μn and μp), and the electron and hole recombination lifetimes (τn and τp) of BiVO4. Me>this sounds podozrive, since they have so many fitting parameters!!!

\paragraph{Model}

\begin{figure}
\includegraphics[width=8.5cm]{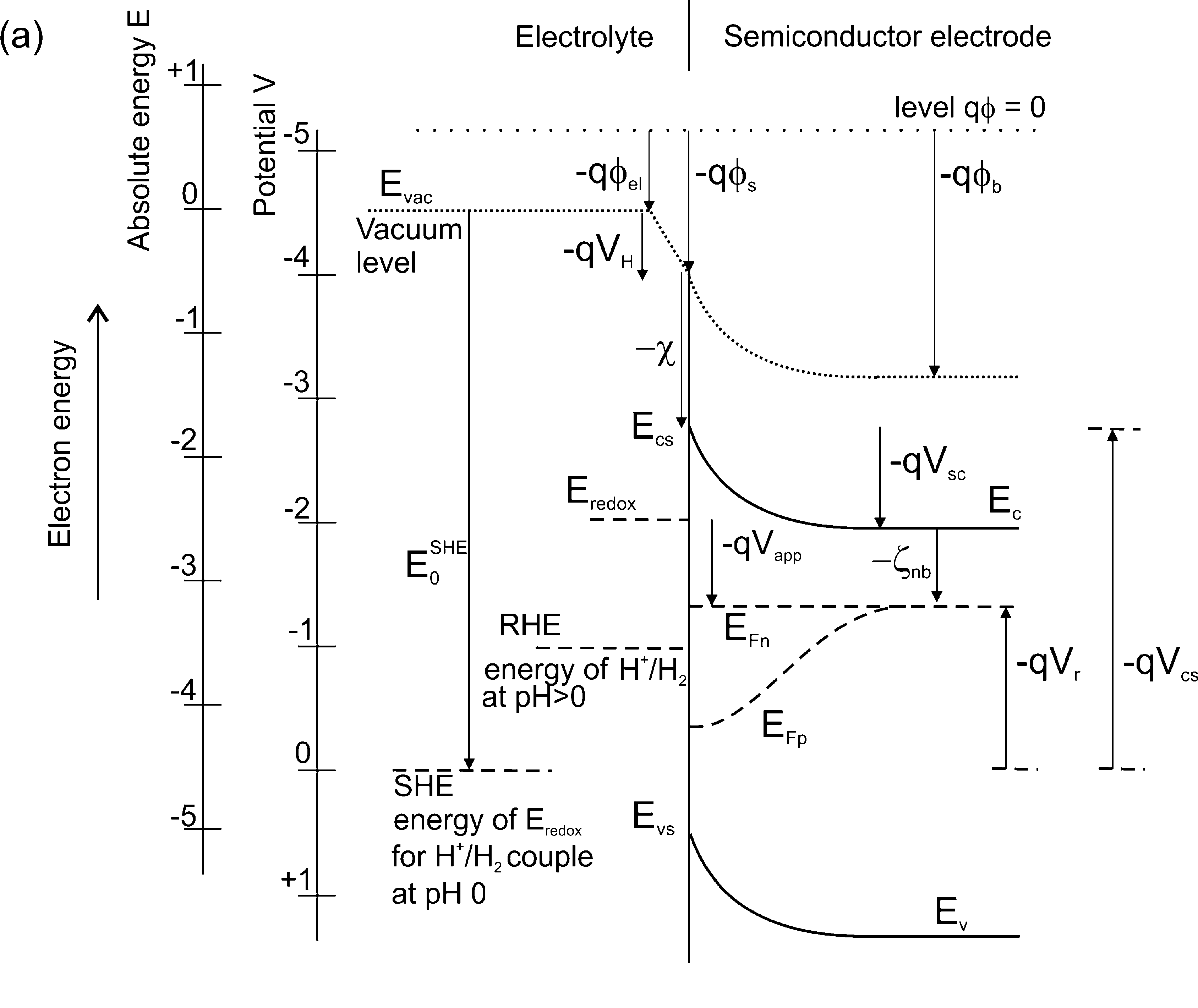}
\caption{Scheme of a n-type semiconductor electrode, with electron energy indicated in the absolute energy scale (with respect to vacuum level), and potentials in the electrochemical scale, with respect to SHE. Reprinted with permission of ACS.}%TODO ask for permission
\label{fig:bisquert_sketch}
\end{figure}

In this work, we present calculation of an energy band diagram of a PEC electrode from a physical model with clearly formulated assumptions \cite{cendula_model_????}. The model is based on charge carrier continuity equations with direct charge transfer from the valence or conduction band to the electrolyte. We consider a PEC cell consisting of an n-type semiconductor with bandgap energy $E_g$, and an electrolyte which can easily accept a single electron or hole (such as H$_2$O$_2$ \cite{dotan_probing_2011} or [Fe(CN)$_6$]$^{3-/4-}$ \cite{klahr_voltage_2011}). Charge transfer occurs across the semiconductor/electrolyte interface until an equilibrium charge distribution is reached and the equilibrium Fermi level in the semiconductor $E_{F0}$ 
becomes equal to the redox Fermi level $E_{redox}$
\begin{equation}
 E_{F0}=E_{redox}.
\end{equation}
We reserve subscript $0$ for equilibrium values in the dark in the following. To derive our model, we use and repeat some of the general definitions introduced in our previous work \cite{bisquert_energy_2013} shown in \ref{fig:bisquert_sketch}. Note that we use notation of subscript $sc$ for semiconductor, $s$ for surface quantity, $b$ for a bulk semiconductor quantity (where electrons and hole remain at equilibrium in the dark). 

Bulk equilibrium properties of the isolated semiconductor are denoted with a subscript $0i$. The bulk of the semiconductor is electrically neutral, hence the concentration of electrons in the bulk $n_{0i}$ must be equal to the number of fully ionized donors $N_D$, $n_{0i}=N_D$. Thus, the concentration of holes is $p_{0i}=n_i^2/n_{0i}$, where $n_i$ denotes intrinsic carrier concentration. An isolated unbiased semiconductor before contact to an electrolyte has a conduction band edge $E_{c,0i}$ and a Fermi level $E_{F,0i}$ related to the vacuum level $E_{vac}$ and electron affinity $\chi$ by %xu schoonen article with table of values. also laser photodetachment spectroscopy. Work function of semiconductor is not so useful, since it gives distance of Fermi level to vacuum, which depends on doping.
%robertson.pdf
\begin{eqnarray}
 E_{c,0i}&=&E_{vac}-\chi,\\
 E_{F,0i}&=&E_{c,0i}-\zeta_{nb},\\
 \zeta_{nb}&=&k_BT\ln\left(\frac{N_C}{n_{0i}}\right), 
\end{eqnarray}
where $k_B$ is the Boltzmann constant, $T$ is the temperature, $q$ is the elementary charge and $N_C$ the effective density of states in the conduction band, and $\zeta_{nb}$ is the distance of the conduction band edge to Fermi level. In the following, we use $E_{vac}$=0 eV as usual. The potential drop in the Helmholtz layer in the dark $V_H$ is calculated from the local vacuum level (LVL) at the surface of the semiconductor ($-q\phi_s$) and LVL of the electrolyte ($-q\phi_{el}$), \ref{fig:bisquert_sketch},
\begin{equation}
 -qV_H=-q\phi_s-(-q\phi_{el}).
\end{equation}
Note that the potential drop in the Helmholtz layer can be a different value at flatband situation (denoted $V_H^{fb}$) than at other measurable voltage (denoted $V_H)$.
 We measure the voltage $V_r$ of the semiconductor electrode with respect to a reference electrode, which means the difference of the Fermi level of electrons in the semiconductor back contact $E_{Fn,b}$ and Fermi level of the reference electrode $E_0^{SHE}$
\begin{equation}  
\label{eq:Vr}
 V_r=-\frac{E_{Fn,b}-E_0^{SHE}}{q}.
\end{equation}
In this article, we use both the Standard Hydrogen Electrode (SHE) energy and Reversible Hydrogen Electrode (RHE) as reference electrodes and scale of energy. Measured voltage with respect to the SHE is denoted $V_r$ (without subscript SHE) and measurable voltage with respect to the RHE $V_{r,RHE}$ with
\begin{equation}
 V_{r,RHE}=V_r+2.3V_{th}\cdot pH,
\end{equation}
where $V_{th}=\frac{k_BT}{q}$ is thermal voltage and $pH$ denotes pH value of the solution. We draw attention to the fact that negative bias versus RHE brings the energy closer to the vacuum level $E_{vac}$. The position of the electron Fermi level at the semiconductor back contact is calculated as (see \ref{fig:bisquert_sketch})
\begin{equation} 
\label{eq:EFnb}
 E_{Fn,b}=-qV_H-\chi-qV_{sc}-\zeta_{nb},
\end{equation}
where $V_{sc}$ denotes the potential drop in the semiconductor. What is usually reported in the literature is the value of flatband potential, which is the measurable voltage when  the bands are flat ($V_{sc}=0$)
\begin{equation} 
\label{eq:Vfb}
 V_{fb}=\left.V_{r}\right|_{V_{sc}=0}=\frac{E_0^{SHE}+\chi+\zeta_{nb}}{q}+V_H^{fb}.
\end{equation}
The value of $V_H^{fb}$ is often not known as it depends on the surface conditions of the semiconductor in the electrolyte. For this article, we know values of $V_{fb}$ and $\chi$ and determine $V_H^{fb}$ from the last equation.  
The potential drop in the semiconductor $V_{sc}$ can be expressed from \ref{fig:bisquert_sketch} as
\begin{equation} 
-qV_{sc}=-q\phi_b-(-q\phi_s).
\end{equation}
Then from eqs.~\ref{eq:Vr},~\ref{eq:EFnb},~\ref{eq:Vfb} follows
\begin{equation} 
V_{sc}=V_r-V_{fb}-(V_H-V_H^{fb}).
\end{equation}
% and it is by some authors \cite{tan_principles_1994} denoted also as the
% built-in voltage $V_{bi}$.%Vbi>0 in tan, archer  book
% Here one has to be careful about sign of energy axis:
% -RHE axis has positive values downwards
% -vacuume axis has positive values upwards
% Thus energy difference will come with different sign for RHE and vacuum axis.

The second option is to refer the voltage to the equilibrium of semiconductor-electrolyte interface (SEI) and this value is denoted $V_{app}$ \cite{bisquert_energy_2013}
\begin{eqnarray}
 V_{app}&=&-\frac{E_{Fn,b}-E_{redox}}{q},\\
 V_{app}&=&V_{sc}-V_{bi}+V_H-V_{H0},
\end{eqnarray}
where built-in voltage is denoted $V_{bi}$ and potential drop across the Helmholtz layer in dark equilibrium $V_{H0}$. Equilibrium of SEI means $V_{app}=0$ V. %TODO comment on value 

%When no oxygen is evolved, $E_{redox}$ in the electrolyte reaches equilibrium with $E_F$. Therefore, position of $E_{redox}$ could be shown to be equal to $E_{F,contact}$ on ~\ref{fig:BandDiagram}.

On the semiconductor side of the junction, the electrostatic potential $\phi$ is obtained by solving Poisson{'}s equation \cite{memming_semiconductor_2008}%also krol_photoelectrochemical_2011
\begin{equation}\label{eq:Poisson}
 \frac{d^2\phi}{dx^2}=-\frac{q(N_D-n(x)+p(x))}{\varepsilon _0\varepsilon _r},
\end{equation}
where $\varepsilon _0$ is the permittivity of vacuum, $\varepsilon_r$ is the relative permittivity of the
semiconductor, $N_D$ is the concentration of fully ionized donors, $n(x)$ is the concentration of free electrons and $p(x)$ is the concentration of free holes ($p(x)\ll n(x)$ for n-type semiconductor in the dark). %krol p.32 and memming p.97
We can write for the conduction and the valence band edge energies $E_{c}$ and $E_{v}$ in the electrostatic potential $\phi(x)$
\begin{eqnarray}\label{eq:BandsSCLJ}
 E_{c}(x)&=&-\chi-q(\phi(x)-\phi_{el}),\\\notag
 E_{v}(x)&=&E_{c}(x)-E_g.
\end{eqnarray}
The band edge pinning (constant value of $E_{c}(0)$ and $E_{v}(0)$) is present if $V_H=V_H^{fb}$ for any measurable voltage (assumed in the following), otherwise the band edges become unpinned.

A simple approximation to solve Poisson{'}s equation, Eq.~\ref{eq:Poisson},
is to assume that the total space charge is uniformly
distributed inside the space charge region (SCR) of width $w$ (also called depletion region approximation)% in schumacher p39, it was introduced by Schottky
\begin{equation}\label{eq:ChargeDis}
w=\sqrt{\frac{2\varepsilon _0\varepsilon _r}{eN_D}\left|V_{sc}\right|}.
\end{equation}
The boundary conditions for the electrostatic potential $\phi$ follow directly from the definitions on \ref{fig:bisquert_sketch}
\begin{eqnarray}\label{eq:PoissonBC}
 \phi(0)&=&\phi_s,\\
 \phi(w)&=&\phi_b.
\end{eqnarray}
The concentration of free electrons and holes in the dark $n_{dark}(x)$ and $p_{dark}(x)$ can be written as 
\begin{eqnarray}\label{eq:ElConcContact}
 n_{dark}(x)&=&n_{0i}\exp^{\frac{\phi(x)-\phi_b}{V_{th}}},\\
 p_{dark}(x)&=&p_{0i}\exp^{\frac{-\phi(x)+\phi_b}{V_{th}}}.
\end{eqnarray}
%The electron concentration at SCR boundary is $n(w)=n_{0i}$ thanks to boundary condition \ref{eq:PoissonBC}. 
%Potential overlaps with lewis 1998.pdf. Compare with energy diagram before contact, after contact in dark, and with illumination in Walter Fig.5. Also tan lewis 1994.pdf
The value of electrostatic potential in the semiconductor bulk $\phi_b$ appears in the above expressions because we have made general definition of electrostatic potential including the potential drop in the Helmholtz layer. Therefore, $\phi_b$ is not zero unlike recent textbook definition \cite{krol_photoelectrochemical_2011}. The approximate solution of Poisson{'}s eq. $\phi_a$ is then 
\begin{eqnarray}\label{eq:ElPotentialParabolic}
\phi_a(x)&=&\phi_b-\text{sign}(V_{sc})\frac{q\, N_D}{2\,\varepsilon_0\,\varepsilon _r}(w -x )^2,
~~~~0<x<w\\\notag
\phi_a(x)&=&\phi_b, ~~~~~~~~~~~~~~~~~~~~~~~~~~~~~~~~w<x<d\,.
\end{eqnarray}
When the measurable voltage $V_r$ is positive of the flatband potential $V_{fb}$, the n-type semiconductor is in the depletion regime. When the measurable voltage is negative of the flatband potential, the semiconductor is in the accumulation regime (due to the sign of $V_{sc}$). 
Upon illumination, we assume low-injection conditions with the number of photogenerated electrons lower than the donor concentration. Thus electron concentration is roughly equal to the dark electron concentration $n(x)=n_{dark}(x)$. The hole continuity equation is solved to obtain free hole concentration $p$ inside of the semiconductor of thickness $d$
\begin{eqnarray}\label{eq:HoleContinuity}%archer&schumacher p.35-36
 0=-\frac{1}{q}\frac{\partial j_h}{\partial x}+G_h(x)-R_h(x).
\end{eqnarray}
We consider the generation rate of charge carriers from the simple Lambert-Beer law $\label{eq:GenRateLambertBeer}G_h(x)=\alpha P\,e^{-\alpha x}$, where $P=\int^{\lambda_g}_{\lambda_{min}} \Phi(\lambda)d\lambda$ is number of photons with energy above $E_g=\frac{hc}{\lambda_g}$ which are absorbed in the semiconductor, $\Phi(\lambda)$ is the spectral photon flux of standard AM15G spectrum with intensity 100 mW/cm$^2$ \cite{nrel_solar_2012}, $\alpha$ is the absorption coefficient of the semiconductor. The hole current density $j_h$ is expressed using the analytical solution of Poisson{'}s equation 
\begin{eqnarray}\label{eq:Fluxes}
 j_h=-qD_h\frac{\partial p}{\partial x}-q\mu_h p\frac{\partial
\phi_a}{\partial x},
\end{eqnarray}
where $\mu_h=\frac{qD_h}{k_BT}$ is the hole mobility, and $D_h$ is the hole diffusion constant. A direct band-to-band nonlinear recombination is assumed 
\begin{equation}
 \label{eq:FirstOrderRecombination}
 R_h=\frac{1}{N_D\tau_h}(n_{dark}p-n_i^2).
\end{equation}

We assume that charge transfer under illumination occurs exclusively from the valence band to the electrolyte. We do not include charge transfer from surface states in the current analysis. The transfer current density of valence band holes at the SEI is described by a first-order approximation  \cite{tan_principles_1994} %p.52(72)
\begin{eqnarray}
\label{eq:FluxBC}
 j_h(0)=-q\;k_{trh}(p(0)-p_{dark}(0))\,, %we could add \exp(-\frac{U_{an}}{k_BT} to include activation energy for anodic process (electron injection from solution to electrode=hole capture from electrode to solution). My problem is that U_{an} is not the same as overpotential so it doesn't simply shift the IV curve to the right. p. 86 morrison book.
\end{eqnarray}
where $k_{trh}$ is the rate constant for hole transfer, and a  linear dependence on the deviation of the interfacial hole concentration $p(0)$ from its dark value $p_{dark}(0)$ at the interface is assumed. Since the thickness of the semiconductor is in the order of the penetration length of light $\alpha^{-1}$ for the hematite parameters listed in \ref{Tab:HematiteParam},  we consider the hole current at the back contact of the semiconductor to depend on a surface recombination velocity $r_s$ 
%TODO find back contact recombination velocity, schlichthoerl minority carrier some estimates
%kroel p.74 frequent TCO properties, which form Ohmic contacts with photoelectrodes
%how it is computed in Comsol for Ohmic contact?
% schumacher p.38 in pdf, [schm]:for Silicon solar cells, good metal contacts have 10 cm/s, bad contacts of bad solar cells 10^6 cm/s. One method for measuring active defects (charge carrier traps) is Deep-level transient spectroscopy (DLTS), used at ISE for Si solar cells.
%gomila 1998.pdf
%little different expression p.43 chuang,
\begin{eqnarray}\label{eq:BCmetalContact}
%j_e(d)=+q\;r_{se}(n(d)-n_{0i}),\\
  j_h(d)=+q\;r_s(p(d)-p_{0i}).
\end{eqnarray}
We use $r_s=10^5$ m/s for numerical calculations throughout this article \cite{foley_analysis_2012}. In order to obtain convergence of the numerical solution procedure, the continuity equation was solved in a non-dimensional form after applying the usual normalization of the variables of the drift-diffusion equations \cite{markowich_semiconductor_1990}.

The quasi-Fermi energies $E_{Fn}, E_{Fp}$ under the influence of an electrostatic potential $\phi(x)$ are calculated by the Boltzmann distribution
\begin{eqnarray}\label{eq:BoltzmannRelations}
 n(x)=N_C\exp\left(-\frac{E_c(x)-E_{Fn}}{k_BT}\right),\\
 p(x)=N_V\exp\left(-\frac{E_{Fp}-E_v(x)}{k_BT}\right).
\end{eqnarray}

\paragraph{Results}
We numerically solved the hole (electron) continuity equation Eq.~\ref{eq:HoleContinuity} for a n-type (p-type) semiconductor by using the depletion region approximation of the electrostatic potential Eq.~\ref{eq:ElPotentialParabolic}. Results for n-type Fe$_2$O$_3$ and p-type Cu$_2$O are presented in the following. If not otherwise stated, we assume $\phi_{el}=0$ V and $V_H^{fb}=V_{H0}=V_H$ in the following.

\begin{figure}
\includegraphics[width=8.5cm]{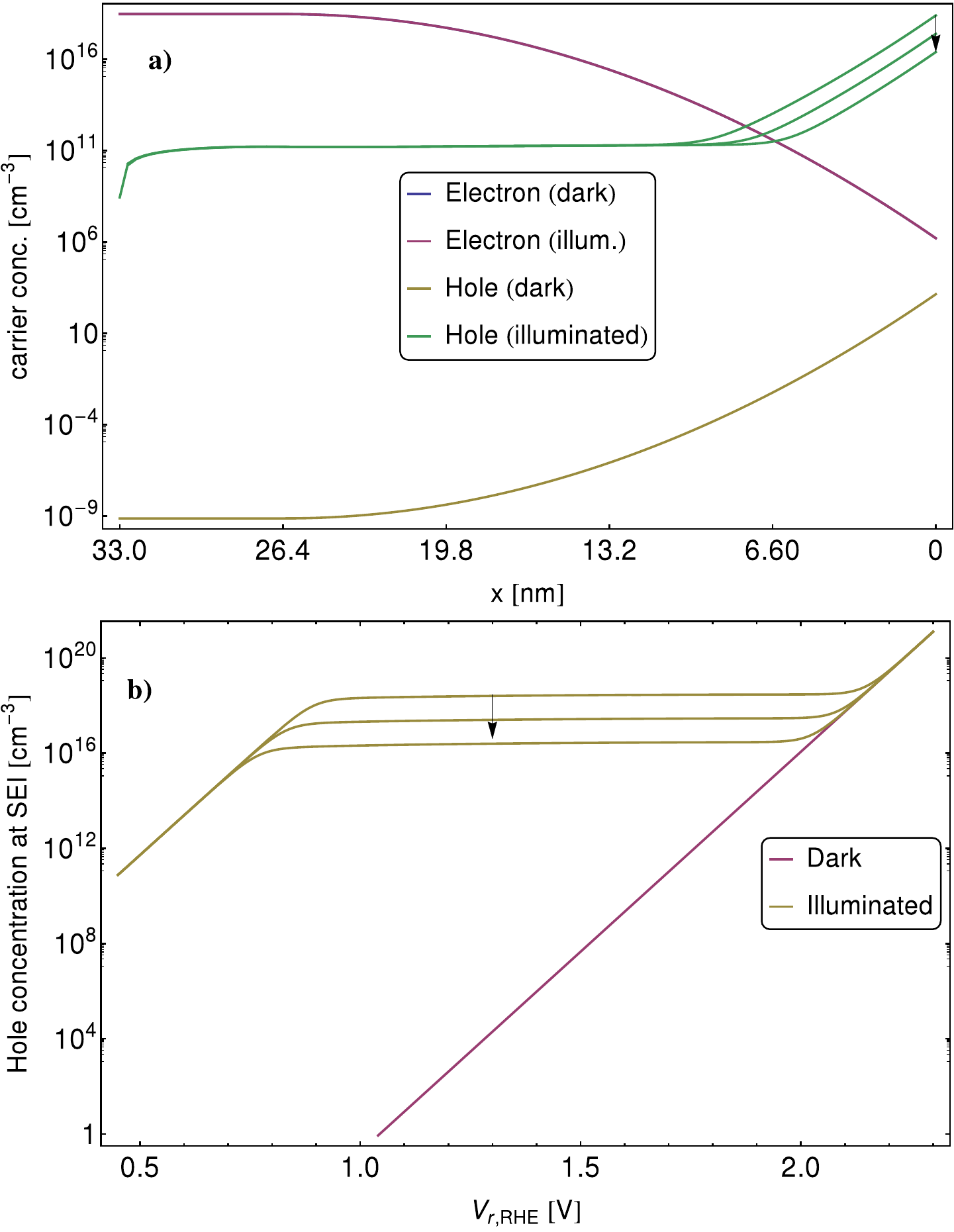}
\caption{a) The simulated charge carrier concentrations in the semiconductor are shown for a measurable voltage of $V_{r,RHE}$=1.23 V. b) Hole concentration at SEI as a function of $V_{r,RHE}$. Direction of arrows mean increasing $k_{trh}$= 10$^{-4}$, 10$^{-3}$, 10$^{-2}$ m/s. Model parameters for hematite from  \ref{Tab:HematiteParam} were used.}
%D.Tilley-hole dark = ionized dopants concentration. Must not be the case since dopants would be much higher concentration?
\label{fig:holeConcIllumPlot}
\end{figure}

\subparagraph{Fe$_2$O$_3$}
The charge carrier concentration profiles calculated from the model are plotted in \ref{fig:holeConcIllumPlot}. In the dark, the SCR is depleted of electrons and the concentrations of holes is increased with respect to the bulk hole concentration. For increasing $V_{r,RHE}$, the dark electron concentration at the SEI $n_{dark}(0)$ decreases until it is smaller than the dark hole concentration at the SEI $p_{dark}(0)$, leading to an inversion layer characterized by a larger concentration of holes (minorities) than electrons (majorities) in the SCR. Corresponding value of $V_{sc}^{inv}=V_{th}\ln\left(\frac{N_D}{n_i}\right)=0.88$ V and thus $V_{r,RHE}^{inv}=1.4$ V are obtained. Therefore, a more detailed future model should take into account the electron continuity equation instead of assuming that the electron concentration upon illumination is equal to the electron concentration in the dark.

Upon illumination, the concentration of electrons is equal to the dark electron concentration. Less holes are accumulated near the SEI for increasing rate constant $k_{trh}$ (faster charge transfer), \ref{fig:holeConcIllumPlot}. For large $V_{r,RHE}$ (>2.0 V), the hole concentration upon illumination near the SEI approaches the hole concentration in the dark, \ref{fig:holeConcIllumPlot}b. At the back contact, the hole concentration follows from solution of continuity equation and boundary condition eq.~\ref{eq:BCmetalContact}. 

\begin{figure}
\includegraphics[width=17cm]{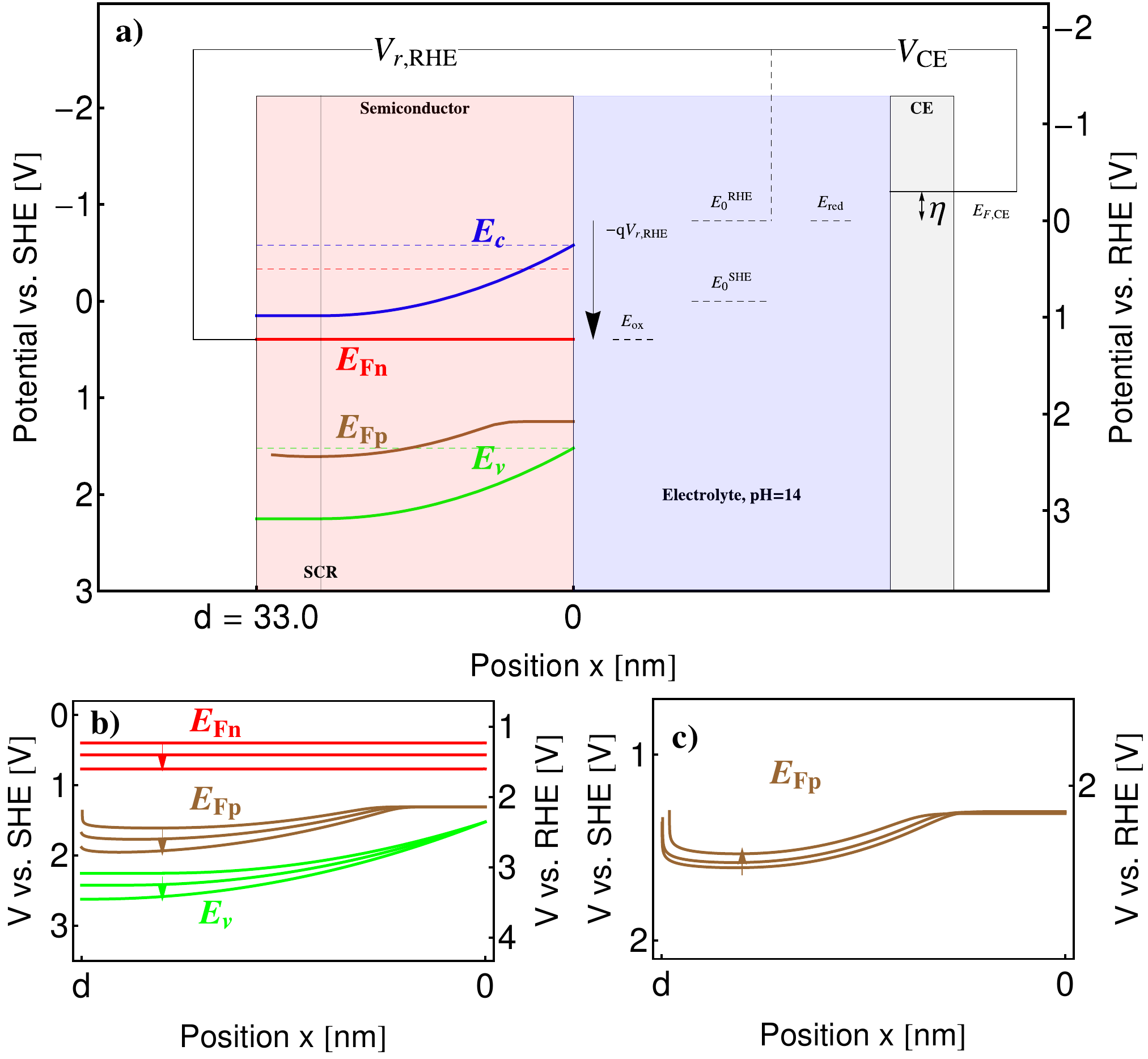}
%\includegraphics[width=17cm]{fig/EnergyBandDiagramPECwaterSplittingDemonstration.pdf}
%D.Tilley-most recombination in SCR where QFL are much closer? Yes confirmed by plotting recombination.
%\includegraphics[width=8.5cm]{../../../EnergyDiagram/export/QFLincreasingVa.pdf}\includegraphics[width=8.5cm]{../../../EnergyDiagram/export/QFLincreasingLh.pdf}
\caption{a) Energy band diagram of n-doped hematite at $V_{r,RHE}$=1.23 V and AM15G sunlight illumination. The semiconductor thickness is $d$, the semiconductor-electrolyte interface is at $x=0\,nm$, and the counter electrode is indicated by CE on the right hand side. A value of $k_{trh}$= 10$^{-3}$ m/s is assumed for the rate constant of charge transfer of valence band holes to the electrolyte. Other material parameters are listed in \ref{Tab:HematiteParam}. An interactive software version to calculate the energy band diagram can be downloaded at \url{http://icp.zhaw.ch/PEC}. b) Quasi-Fermi level diagram, where the arrows indicate increasing values of the measurable voltage, $V_{r,RHE}$=1.23, 1.4, 1.6 V. c) Influence of the minority carrier diffusion length $L_h$ on the Quasi-Fermi level $E_{F_p}$ of holes, where the arrow indicates increasing values of $L_h$=5, 10, 25 nm.}
\label{fig:BandDiagram}
\end{figure}

The energy band diagram is shown for a three-electrode setup in \ref{fig:BandDiagram}. A  measurable voltage of $V_{r,RHE}=1.23$ V is assumed, which is the standard voltage used for comparison of different PEC electrodes \cite{kay_new_2006,chen_efficiency_2013}. The measurable voltage $V_{r,RHE}$ is indicated in \ref{fig:BandDiagram}a) with an arrow on the energy scale, $-qV_{r,RHE}$. This is also explained in our previous work \cite{bisquert_energy_2013}. The band edges of the semiconductor $E_c(x),E_v(x)$ for flatband condition ($V_{r,RHE}=V_{fb,RHE}$) are shown as dashed lines, whereas those away from flatband condition ($V_{r,RHE}\neq V_{fb,RHE}$) are shown as solid lines. Band positions at flatband conditions for hematite agree well with values reported for example for $pH=1$ \cite{nozik_photoelectrochemistry:_1978,gratzel_photoelectrochemical_2001} and $pH=14$ \cite{krol_solar_2008}.  An upward band bending of the semiconductor is present if $V_{r,RHE}$ V is more positive than $V_{fb,RHE}$, see \ref{
fig:BandDiagram}. Band edges are pinned at the SEI by default (since we assume $V_H=V_H^{fb}$), but we allow for modification of surface conditions by changing the value of $V_H$ in our interactive band diagram software \cite{cendula_model_????}.

% Simple TESTs for energy band diagram to agree with band diagrams in the literature \cite{tan_principles_1994,Fig. 2.17 p. 41 krol }.
% \begin{itemize}
%  \item $V_a<Vfb$ no depletion layer can form -> no photocurrent
%  \item $V_a=Vfb$ flat-bands are observed 
%  \item $V_a>Vfb$ should start and increase band bending (depletion layer is formed) -> charges can be separated and photocurrent onsets
% \end{itemize}
%also nozik memming fig.9-10
%also \cite{bak_photo-electrochemical_2002} fig. 4-7

% \begin{itemize}
%  \item position of $E_{C0}$ compared to krol fig. 3
%  \item shift of $E_{C0}$ with applied potential
%  \item band positions stay the same at the interface \cite{bard_electrochemical_1980} p.768
%  \item $E_{CB}$ decreases in SCR
% \end{itemize}

% Other sources of band diagrams for hematite:

% andrade transient ph=13.6?
% gratzel for ph=1 in Fig.2, referenced in Hu Improved 2009.pdf
% krol book p. 130 for ph=13.6 (1M NaOH)
% sivula review Fig.1 

The number of photogenerated electrons is small compared to the donor concentration, and thus illumination does not change the electron concentration. Therefore, the electron quasi-Fermi level $E_{Fn}$ is also constant across the semiconductor, eq. \ref{eq:BoltzmannRelations}, and $E_{Fn}=E_{Fn,b}$. The position of $E_{Fn}$ relative to $E_0^{RHE}$ in the energy diagram is given by arrow $-qV_{r,RHE}$, eq. \ref{eq:EFnb}. In contrast, the hole concentration is determined mainly by photogenerated holes, which are redistributed in the semiconductor according to the continuity equation Eq.~\ref{eq:HoleContinuity}. Since $E_{v}(0)$ is more positive than $E_{ox}$, a transfer of holes from the valence band can thermodynamically oxidize the electrolyte species. An external wire electrically connects the semiconductor to the metal counter electrode (CE) through a potentiostat. The counter electrode Fermi level $E_{F,CE}$  is automatically adjusted by 
applying a voltage $V_{CE}$ above the water reduction energy $E_{red}$ (including the electrochemical overpotential $\eta$) by the potentiostat to enable hydrogen evolution at the counter electrode. The counter electrode is shown in the energy diagram only to completely describe a three-electrode setup and we ignore its polarization in the following \cite{hodes_photoelectrochemical_2012}. In the electrolyte, we plot two reference electrode energies $E_0^{SHE}$ and $E_0^{RHE}$, standard water reduction and oxidation energy $E_{red}$(0 eV vs RHE) and $E_{ox}$(1.23 eV vs RHE). Note that the relation of $E_{red}$ and $E_{ox}$ to $E_{redox}$ depends on the concentrations (activities) of oxidizing and reducing species in the solution \cite{morrison_electrochemistry_1980}. %under conditions of vigorous oxygen-evolution at the photoanode we can say that Eredox =Eox= +1.23V vs. RHE. At the counter electrode, again under conditions of vigorous hydrogen 
%evolution, Eredox = Ered=0V vs. RHE.[D.Tilley]

The energy band diagram in the semiconductor for different values of the measurable voltage $V_{r,RHE}$ is plotted in \ref{fig:BandDiagram}b). For increasing $V_{r,RHE}$ the band bending increases and the electron quasi-Fermi level $E_{Fn}$ shifts down on the RHE scale. Interestingly, the hole quasi-Fermi level  $E_{Fp}(0)$ at the SEI remains nearly constant for increasing $V_{r,RHE}$ (see Figure S1 in Supporting Information) and thus the splitting of the quasi-Fermi levels (photovoltage) approaches zero.  In the neutral region $w<x<d$, the hole quasi-Fermi level $E_{Fp}(x)$ is more negative for increasing $V_{r,RHE}$ and the photovoltage is nearly constant. When the hole diffusion length $L_h=\sqrt{D_h \tau_h}$ is increased, the flat region of the hole quasi-Fermi level $E_{Fp}$ near the SEI is enlarged, \ref{fig:BandDiagram}c), and the hole concentration in the neutral region decreases (see Figure S2 in Supporting Information). 

% It is useful to define splitting of quasi-Fermi level as photovoltage%, corresponding to the maximum chemical energy of the electron-hole pair
%  \begin{equation}
%  V_{ph}(x)=E_{Fn}(x)-E_{Fp}(x). 
%  \end{equation}
%THIS IS NOT GOOD DEFINITION. Rather V_{ph}=E_{Fn}-E_{redox} see memming p. 41, fig 2.12a

\begin{figure}
\includegraphics[width=13cm]{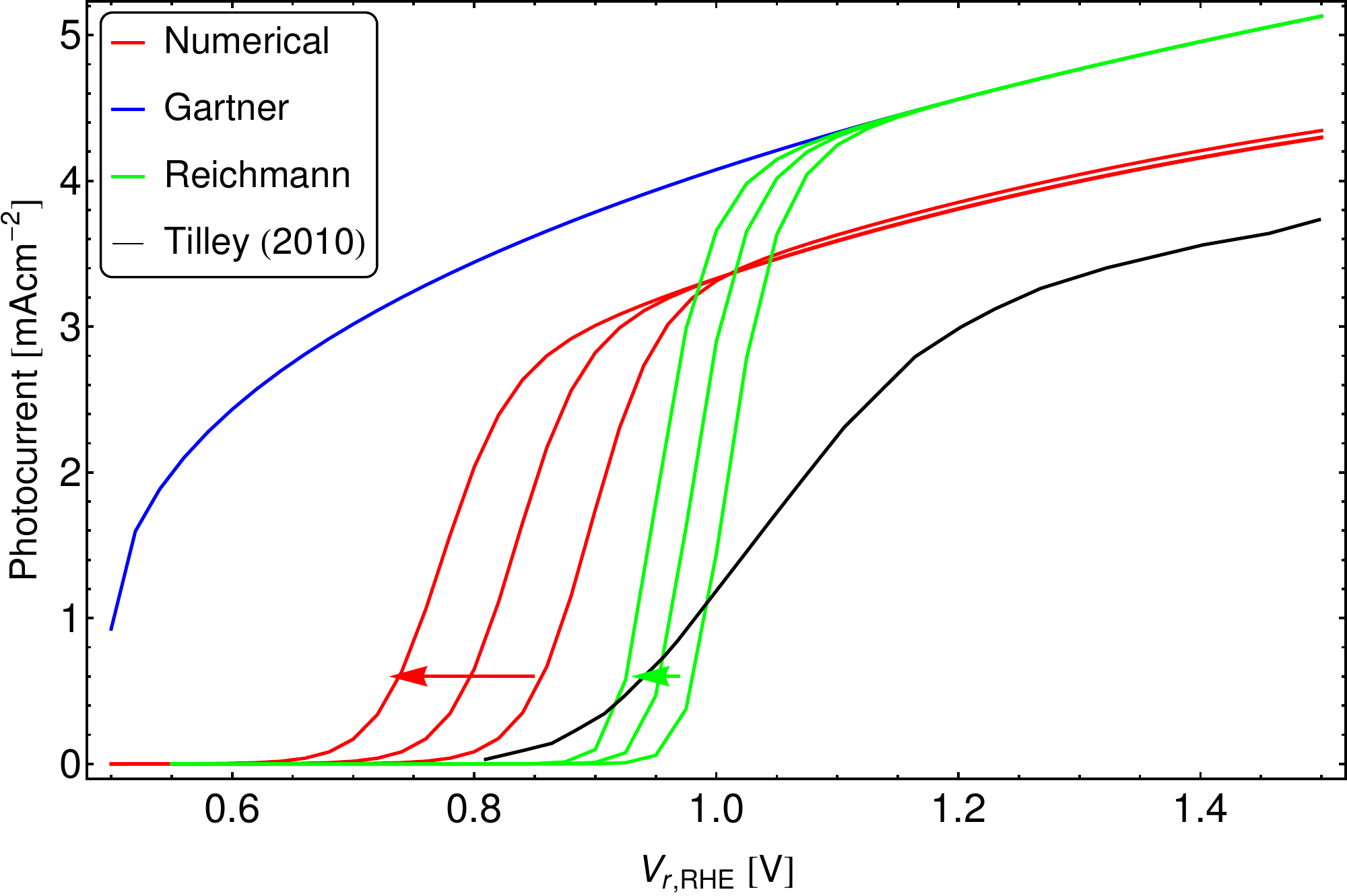}
\caption{Current-voltage curves for $k_{trh}$= 10$^{-4}$, 10$^{-3}$, 10$^{-2}$ m/s (in direction of arrows) from our numerical model, Gartner model and Reichmann model for n-doped hematite and other material parameters listed in \ref{Tab:HematiteParam}.}
\label{fig:IVcurve}
\end{figure}

%I don't have to include the j_e(0) in my total current because: exchange electron current density is small and diode equation doesn't amplify it to scale comparable to j_h (tan_principles_1994 : illuminated I-V characteristic)
We simulated current-voltage curves with our numerical model $j_h(0)$ (eq. \ref{eq:FluxBC}) and compared the results with published models of Gartner \cite{gartner_depletion-layer_1959} and Reichmann \cite{reichman_currentvoltage_1980}, \ref{fig:IVcurve}. According to Gartner, the minority charge carrier concentration is calculated from the diffusion equation, neglecting recombination in the SCR and assuming that every hole in SCR contributes to the photocurrent. Photocurrent density of Gartner is
\begin{equation}\label{eq:gartner} %eq from cass phd
 j_G=eP\left(1-\frac{e^{-\alpha w}}{1+\alpha L_h}\right).
\end{equation}
Therefore, $j_G$ overestimates the minority carrier photocurrent in comparison to our numerical model $j_h(0)$. Recombination in the SCR by Sah-Noyce-Shockley formalism was incorporated into the model by Reichmann \cite{reichman_currentvoltage_1980} with resulting photocurrent $j_R$ (the detailed expression is given in the Supporting Information). For small $V_{r,RHE}$, $j_R$ is much smaller than $j_G$ because the SCR recombination loss is included in $j_R$. The onset of the photocurrent calculated by Reichmann $j_R$ starts when $\gamma=\frac{j_se^{-\frac{V_{app}}{V_{th}}}}{qk_{trh}p_{dark}(0)}\approx 1$ ($j_s$ is the saturation current density as defined in the SI). Therefore, if we consider faster charge transfer kinetics (larger $k_{trh}$), we need a 
smaller value of the onset potential $V_{r,RHE}$ (and thus $V_{app}$) to obtain a similar value $\gamma\approx1$. For increasing  $V_
{r,RHE}$, $j_R$ approaches $j_G$ because the SCR recombination becomes negligible in $j_R$ \cite{reichman_currentvoltage_1980}, but the numerical photocurrent $j_h(0)$ is smaller than $j_G$ since SCR recombination is included in $j_h(0)$. The numerical photocurrent $j_h(0)$ onsets when $V_{r,RHE}$ is more positive than $V_{fb,RHE}$ and it is larger than $j_R$ for small $V_{r,RHE}$.  %, since our band-to-band recombination model (eq. \ref{eq:FirstOrderRecombination}) is not so effective as SRH recombination included in $j_R$. unfinished -plot of recombination rate as a function of $V_{r,RHE}$ 
 Increasing the rate constant $k_{trh}$ represents a faster exchange rate of holes with the solution. This also shifts the numerical j-V curve  to the left as predicted by the Reichmann model, decreasing the onset potential of the photocurrent.%TODO one might extract relation between ktrh and onset potential.
  
  Measured current-voltage curve for nanostructured Fe$_2$O$_3$ electrode with IrO$_2$ catalyst \cite{tilley_lightinduced_2010} is compared with the prediction from our model on \ref{fig:IVcurve}. The onset voltage of measured photocurrent $\approx$ 0.8 V$_{RHE}$ is reproduced with numerical photocurrent with $k_{trh}$=  10$^{-4}$ m/s. However, the slope of measured photocurrent and its value 4.3 mA/cm$^2$ at 1.5 V$_{r,RHE}$ is smaller than the slope of numerical photocurrent and its value 3.8 mA/cm$^2$ at 1.5 V$_{r,RHE}$. These differences point out that we cannot verify our model description by comparing the current-voltage curves alone, because the kinetic effects cannot be distinguished in the current-voltage response. Comparison of the model predictions with impedance spectroscopy \cite{klahr_electrochemical_2012}  measurements is needed to verify the model. 
 
We checked that the maximum photocurrent obtainable from the hematite electrode based purely on the number of absorbed photons is $qP=12.5$ mA/cm$^2$ for AM15G illumination. This value is also obtained for the Gartner photocurrent eq.~\ref{eq:gartner} when the bracket term is close to one and also for the Reichmann photocurrent (which recovers the Gartner photocurrent in regime of large voltages). The plateau of numerical photocurrent $j_h(0)$ cannot be computed here, because our model cannot be used to predict photocurrents at voltages higher than $V_{r,RHE}>V_{r,RHE}^{inv}$. At such voltages inversion layer is formed as described in the previous text and this would need degenerate statistics to be included in the model.

\begin{figure}
\includegraphics[width=17cm]{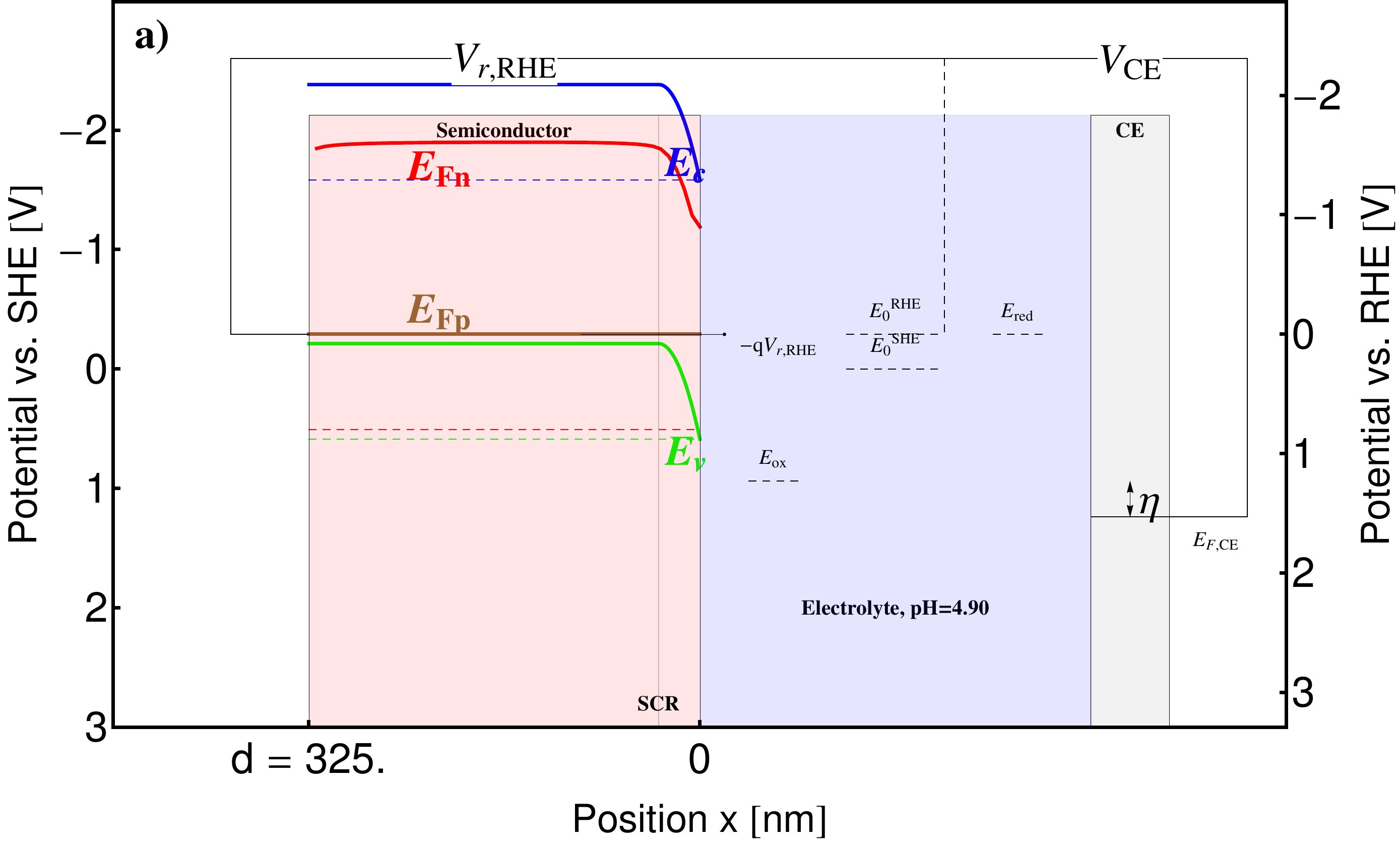}
\caption{Energy band diagram for p-doped Cu$_2$O, $k_{tre}$= 10 m/s and $V_{r,RHE}$=0 V (hence the voltage arrow is not visible in the diagram). The  material parameters are listed in  \ref{Tab:HematiteParam}. The interactive version of this figure can be accessed at \url{http://icp.zhaw.ch/PEC}.}
\label{fig:BandDiagramCu20}
\end{figure}

\subparagraph{Cu$_2$O}
We also applied our model to simulate p-type semiconductors for photocathodes. Appropriate changes in the equations were introduced, resulting from doping with acceptors rather than donors. Cuprous oxide (Cu$_2$O) is an abundant and promising material for PEC photocathodes. The main issue with Cu$_2$O is its limited stability in water, which is currently being addressed with stabilizing overlayers \cite{paracchino_highly_2011,paracchino_ultrathin_2012,tilley_ruthenium_2013}. A downward band-bending occurs when $V_{r,RHE}$ is more negative than $V_{fb,RHE}$. This leads to a drift of electrons to the electrolyte, \ref{fig:BandDiagramCu20}. Upon illumination, the hole concentration is assumed to remain equal to the dark hole concentration. The electron concentration is calculated from the electron continuity equation. Electrons are accumulated near the SEI where they reduce water to H$_2$ with a rate constant $k_{tre}$. 

In the case of p-type Cu$_2$O, the majority carriers are holes, and thus the counter electrode carries out the oxidation reaction (including the associated overpotential $\eta$). Although the electron quasi-Fermi level $E_{Fn}$ is negative with respect to $E_{red}$ making it suitable for hydrogen evolution, \ref{fig:BandDiagramCu20}, corrosion prevents hydrogen evolution in the experiment unless the Cu$_2$O is protected by overlayers \cite{paracchino_highly_2011}. So far, our model does not consider corrosion; here we aimed at showing the general energetic configuration of a p-type PEC photoelectrode. 

%Do not comment on IV curves for Cu2O or BiVO4, since I don't show them explicitly in the article. The onset potential in our simulation (roughly V_{f b} does not relate well with the measured onset potential - 0.55V_RHE for Cu2O \cite{tilley_ruthenium_2013} and 0.6 V_RHE for BiVO4 \cite{abdi_efficient_2013}

%\cite{paracchino_highly_2011}: onset potential around 0.35 V_RHE, looks quite linear
%my energy diagram: onset potential close to Vfb=0.8 V_RHE
%memming p.176 : large difference between IV curves for n- and p-type semiconductors
%de Jongh: j-V curves look worse than paracchino

% \begin{figure}
% \includegraphics[width=13cm]{../../../C228/Cu2O/CarrierConc.png}
% \includegraphics[width=13cm]{../../../C228/Cu2O/QFLVa0.png}
% \caption{P-type semiconductor Cu$_2$O. Carrier concentrations and energy band diagram for $V_{r,RHE}$=0 V$_{RHE}$ numerically computed with recombination in SCR. Material parameters from \ref{Tab:HematiteParam} are used.}
% \label{fig:BandDiagramComsolVa07}
% \end{figure}

%TODO access to demonstration only for registered users to keep record of who used the software.
%Creative Commons Attribution—NonCommercial—ShareAlike 3.0, put logo and link in home.zhaw.ch

%TODO Note that there is no need to know value of $V_{bi}$, since it gives values of $E_{redox}=E_{F,0i}+qV_{bi}$ and enables definition of $V_{app}=V_{sc}-V_{bi}+(V_H-V_{H0})$ according to Bisquert \cite{bisquert}.

\paragraph{Conclusion}
We presented a physical model for minority charge carrier transport in semiconductor PEC electrodes in contact with an electrolyte. Direct charge transfer to the electrolyte from valence or conduction band, band-to-band recombination and Lambert-Beer optical generation were assumed. A numerical solution of the model equations allows us to calculate the minority carrier concentration and the quasi-Fermi level. A resulting energy band diagram of a PEC cell accounts for the potential drop in the Helmholtz layer and it is capable of modeling both band edge pinning and unpinning. Comparison of the numerically obtained photocurrent with analytical results and measurement reveals need for verification of the model with spectroscopic measurements. Our model was implemented in an interactive software that can be freely accessed online \cite{cendula_model_????}. All presented results of this article can be reproduced with 
this software and we invite all members of the research community to use it while designing PEC cells. We are currently working on an extension of our model to a fully coupled  drift-diffusion model with surface states. Such photoelectrode models need to accompany the experimental studies to suppress recombination losses (e.g. by surface passivation) and enhance charge transfer (e.g. by catalysis), the two major issues for efficient metal oxide photoelectrodes \cite{sivula_metal_2013}. %Including surface states will enable an accurate physical description of PEC electrodes similar to the established simulations tools for the electronics industry \cite{markowich_semiconductor_1990}.

%We have additionally included n-type BiVO$_4$ in our software as another promising candidate for PEC electrodes. 
% The current through the SEI fits well to the linear j-V behavior observed for hematite in one-electron redox couple [Fe(CN)$_6$]$^{3-/4-}$ \cite{klahr_voltage_2011}. The extracted charge transfer rate constant shows a linear increase with light intensities below 1800 W/m$^2$. A roughly constant value is found for illumination 
% above this intensity.

% \begin{figure}
% \includegraphics[width=13cm]{../../../C228/QFLVa07.png}
% \includegraphics[width=13cm]{../../../C228/IV.png}
% \caption{Energy band diagram at $V_{r,RHE}$=0.7 V$_{RHE}$ numerically computed with recombination in SCR. Colors of the individual energy bands correspond to \ref{fig:BandDiagram}.  Hematite parameters from \ref{Tab:HematiteParam} are used. Unfortunately, it's not possible at current stage to include scale of RHE and SHE in the plot instead of vacuum scale.}
% \label{fig:BandDiagramComsolVa07}
% \end{figure}

%\input{OpticalSimulations}


\begin{acknowledgement}

We thank F.T. Abdi, H. J. Lewerenz, G. Schlichthoerl and B. Klahr for fruitful discussions. Financial support by the Swiss Federal Office of Energy (PECHouse2 project, contract number SI/500090-02) is gratefully acknowleged.

\end{acknowledgement}


\begin{table}[h]
\begin{tabular}{|l|l|l|l|l|}
\hline
Symbol                                 &    Fe$_2$O$_3$\cite{upul_wijayantha_kinetics_2011} & Cu$_2$O \cite{paracchino_ultrathin_2012,paracchino_synthesis_2012}     & Description  	\\ \hline
$N_D$ [cm$^{-3}$]        &    $2.91\cdot10^{18}$  & 0 & Donor concentration\\
$N_A$ [cm$^{-3}$]        &    0 &  $5\cdot10^{17}$  &Acceptor concentration\\ 
$V_{fb,RHE}$ [V]      &    +0.5 %p. 95 
& +0.8 &  Flatband potential \\
$\chi$ [eV]      &    +4.78\cite{xu_absolute_2000,niu_hydrothermal_2010}
& +4.22 \cite{xu_absolute_2000} &  Electron affinity \\
$N_C$ [cm$^{-3}$         & $4\cdot10^{22}$ \cite{morin_electrical_1954,cesar_influence_2008}	&  $1.1\cdot10^{19}$ 
& Density of states of CB  \\%measurement technique which could be applied to hematite as well, enright 1996.pdf
$N_V$ [cm$^{-3}$]			& $1\cdot10^{22}$ &	$1.1\cdot10^{19}$   & Density of states of VB \\
$\varepsilon_r$               & 32 \cite{glasscock_structural_2008} & 6.6 &Relative permitivity\\
$E_g$ [eV]                          &  2.1 & 2.17	& Bandgap energy\\
$d$ [nm]				& 33 & 325 & Thickness of semiconductor \\
$\tau_{e}$ [ns]                          & -    & 0.25 & Electron lifetime\\
$\tau_{h}$ [ns]                          & 0.048\cite{bosman_small-polaron_1970} & -%0.19 
&Hole lifetime\\
$L_{e}$ [nm]                          & -    & 40 & Electron diffusion length\\
$L_{h}$ [nm]                          &    5 \cite{krol_photoelectrochemical_2011}& -%50 
& Hole diffusion length\\
$\alpha$ [cm$^{-1}$]                  & $1.5\cdot10^{5}$ & $1.3\cdot10^{4}$ & Absorption coefficient\\
%\hline
$pH$                      &    14 & 4.9 & pH value of the electrolyte \\
%$I$ [Wm$^{-2}$]                       & 1000  & 1000&  Illumination power density\\
%$\lambda$ [nm]                        & 455  & 455 & Wavelength of illumination \\
\hline
\end{tabular}
\caption{Material parameters of  semiconductors used in the calculations.}
\label{Tab:HematiteParam}
\end{table}

% \begin{figure}
% \includegraphics[width=17cm]{../../../EnergyDiagram/export/klahrAPLFig2purple.pdf}
% \includegraphics[width=17cm]{../../../EnergyDiagram/export/ktrVsIntensity.pdf}
% \caption{a) Comparison of measured (points, \cite{klahr_voltage_2011}) and fitted (solid lines) j-V curves for various light intensities. b) Dependence of extracted values of $k_{tr}$ from a) on the light intensity. The error bars are shown as well. Dashed lines present trends of the data below and above an illumination intensity of 1800 W/m$^2$. The parameters are listed in \ref{Tab:HematiteParam}. 
% }
% \label{fig:klahrAPLFig2purple}
% \end{figure}

% \begin{figure}
% \includegraphics[width=17cm]{fig/holeConcIllumPlotCu2O.pdf}
% \caption{Carrier concentrations upon illumination in the p-type Cu$_2$O from the numerical model for $V_{r,RHE}$=1.23 V$_{RHE}$ and parameters from \ref{Tab:HematiteParam}.}
% \label{fig:holeConcIllumPlot}
% \end{figure}

\begin{table}[h]\footnotesize
\begin{tabular}{|l|l|l|}

\hline
Symbol          & Unit &  Description \\
\hline
CE & & Counter electrode\\ 
LVL & & Local vacuum level\\ 
PEC & & Photoelectrochemical\\
SCR & & Space-charge region\\
SEI & & Semiconductor-electrolyte interface\\
SHE & & Standard hydrogen electrode\\
SI  & & Supporting information\\
RHE & & Reversible hydrogen electrode\\

\textit{Subscript i} & & Quantity in the isolated semiconductor \ before contact to an electrolyte\\
\textit{Subscript b} & & Quantity in the semiconductor bulk\\
\textit{Subscript s} & & Quantity at the SEI\\

$k_{B}$         & eV/K&   Boltzmann constant ($8.6\cdot10^{-5}$ eV/K)\\
$T$         & K&   Temperature (300 K)\\
$q$         & C&   Elementary charge ($1.6\cdot 10^{-19}$ C)\\
$V_{th}$    & V&   Thermal voltage (25.9 mV)\\
$h$         & J$\cdot$s&   Planck's constant ($6.62607\cdot10^{-34}$ J$\cdot$s)\\
$c$         & m/s&   Speed of light in vacuum (299792458 m/s)\\

$V_{r}$         & V&   Measurable voltage with respect to SHE reference electrode\\
$V_{r,RHE}$     & V&    Measurable voltage with respect to RHE\\
$V_{r,RHE}^{inv}$     & V&    Measurable voltage with respect to RHE when the inversion layer starts to form\\
$V_{fb}$     & V&    Flatband voltage with respect to SHE\\
$V_{fb,RHE}$     & V&    Flatband voltage with respect to RHE\\
$V_{app}$       & V&    Applied voltage to the semiconductor with respect to the dark equilibrium (unbiased)\\
$V_{H}$     & V&    Potential (voltage) drop across the Helmholtz layer in the dark\\
$V_{H}^{fb}$     & V&    Potential (voltage) drop across the Helmholtz layer at flatband situation in the dark\\
$V_{H0}$     & V&    Potential (voltage) drop across the Helmholtz layer in the dark equilibrium\\
$V_{sc}$     & V&    Potential (voltage) drop across the semiconductor\\
$V_{bi}$     & V&    Built-in voltage of semiconductor/liquid junction\\
$V_{CE}$     & V&    Voltage between the reference electrode and counterelectrode\\
$\eta$   & V &   Electrochemical overpotential at the CE  \\ %(containing both kinetic and activation overpotentials)

$E_{vac}$     & eV&    Energy of the local vacuum level\\
$E_0^{SHE}$     & eV&    Energy of the SHE with respect to vacuum level of the electron (-4.44 eV)\\
$E_0^{RHE}$     & eV&    Energy of the RHE with respect to vacuum level of the electron\\
$E_{redox}$     & eV&    Fermi level of the electrolyte species (redox level)\\
$E_{red}$     & eV&     Standard water reduction energy\\
$E_{ox}$     & eV&    Standard water oxidation energy\\
$E_{F0}$     & eV&    Equilibrium Fermi level in the semiconductor (dark)\\
$E_{c,0i}$     & eV&    Conduction band edge in the isolated semiconductor before contact to an electrolyte\\
$E_{F,0i}$      & eV&    Fermi level in the isolated semiconductor before contact to an electrolyte\\
$E_{Fn}$, $E_{Fp}$      & eV&   Quasi-Fermi energy of electrons and holes\\
$E_{Fn,b}$      & eV&    Quasi-Fermi energy of electrons at the back contact\\
$E_{c}$     & eV&    Conduction band edge in the semiconductor\\
$E_{v}$     & eV&    Valence band edge in the semiconductor\\
$E_{F,CE}$   & eV &   Fermi level of the CE \\
$\zeta_{nb}$      & eV&    The difference between the semiconductor conduction band energy and the electron Fermi level\\

$\phi$      & V&    Local electrostatic potential\\
$\phi_a$      & V&     Approximate solution for local electrostatic potential\\
$\phi_{el}$      & V&    Local electrostatic potential of the electrolyte\\
$\phi_{s}$      & V&    Local electrostatic potential at SEI\\
$\phi_{b}$      & V&    Local electrostatic potential in the semiconductor bulk\\

\hline
\end{tabular}
\caption{Table of symbols and abbreviations. Symbols for material parameters are defined in \ref{Tab:HematiteParam}.}
\label{Tab:TOS}
\end{table}

\begin{table}[h]\footnotesize
\textbf{Table 2 continuted.}
\begin{tabular}{|l|l|l|}
\hline
Symbol          & Unit &  Description \\
\hline
$n_{i}$          & m$^{-3}$&   Intrinsic carrier concentration in the bulk of semiconductor \\
$n_{0i}$,$p_{0i}$ & m$^{-3}$&   Equilibrium concentration of electrons and holes in the bulk of isolated semiconductor \\
$n_{dark}$, $p_{dark}$         & m$^{-3}$&   Dark concentration of electrons and holes \\
$n$, $p$         & m$^{-3}$&   Concentration of electrons and holes \\
$w$         & m&   Width of the space-charge region in the semiconductor \\
$j_h$         & A/m$^2$&   Hole current density \\
$j_G$         & A/m$^2$&   Current density calculated by Gartner \cite{gartner_depletion-layer_1959} \\
$j_R$         & A/m$^2$&   Current density calculated by Reichmann\cite{reichman_currentvoltage_1980}\\
$j_s$         & A/m$^2$&   Saturation current density \\
$G_h$,$R_h$   & m$^{-3}$s$^{-1}$&   Generation and recombination rate of holes \\
$P$         & m$^{-2}$s$^{-1}$&   Number of photons absorbed in the semiconductor from AM15G spectrum \\
$\Phi$         & m$^{-3}$s$^{-1}$&   Spectral photon flux of AM15G spectrum \\
$\mu_h$   & m$^{2}$V$^{-1}$s$^{-1}$ &   Mobility of holes \\
$D_h$   & m$^{2}$s$^{-1}$ &   Diffusion constant of holes \\
$k_{trh}$   & ms$^{-1}$ &   Rate constant for charge transfer of VB holes to electrolyte \\
$\lambda_g$ & m &   Wavelength below which semiconductor absorbs photons \\
$r_s$   & ms$^{-1}$ &   Back contact surface recombination velocity \\

\hline
\end{tabular}

\label{Tab:TOS2}
\end{table}

\bibliography{../../../report/report.bib}

\providecommand*{\mcitethebibliography}{\thebibliography}
\csname @ifundefined\endcsname{endmcitethebibliography}
{\let\endmcitethebibliography\endthebibliography}{}
\begin{mcitethebibliography}{48}
\providecommand*{\natexlab}[1]{#1}
\providecommand*{\mciteSetBstSublistMode}[1]{}
\providecommand*{\mciteSetBstMaxWidthForm}[2]{}
\providecommand*{\mciteBstWouldAddEndPuncttrue}
  {\def\EndOfBibitem{\unskip.}}
\providecommand*{\mciteBstWouldAddEndPunctfalse}
  {\let\EndOfBibitem\relax}
\providecommand*{\mciteSetBstMidEndSepPunct}[3]{}
\providecommand*{\mciteSetBstSublistLabelBeginEnd}[3]{}
\providecommand*{\EndOfBibitem}{}
\mciteSetBstSublistMode{f}
\mciteSetBstMaxWidthForm{subitem}{(\alph{mcitesubitemcount})}
\mciteSetBstSublistLabelBeginEnd{\mcitemaxwidthsubitemform\space}
{\relax}{\relax}

\bibitem[Lewis and Nocera(2006)]{lewis_powering_2006}
Lewis,~N.~S.; Nocera,~D.~G. \emph{Proceedings of the National Academy of
  Sciences} \textbf{2006}, \emph{103}, 15729--15735\relax
\mciteBstWouldAddEndPuncttrue
\mciteSetBstMidEndSepPunct{\mcitedefaultmidpunct}
{\mcitedefaultendpunct}{\mcitedefaultseppunct}\relax
\EndOfBibitem
\bibitem[Schiermeier(2013)]{schiermeier_renewable_2013}
Schiermeier,~Q. \emph{Nature} \textbf{2013}, \emph{496}, 156--158\relax
\mciteBstWouldAddEndPuncttrue
\mciteSetBstMidEndSepPunct{\mcitedefaultmidpunct}
{\mcitedefaultendpunct}{\mcitedefaultseppunct}\relax
\EndOfBibitem
\bibitem[Khaselev and Turner(1998)]{khaselev_monolithic_1998}
Khaselev,~O.; Turner,~J.~A. \emph{Science} \textbf{1998}, \emph{280},
  425--427\relax
\mciteBstWouldAddEndPuncttrue
\mciteSetBstMidEndSepPunct{\mcitedefaultmidpunct}
{\mcitedefaultendpunct}{\mcitedefaultseppunct}\relax
\EndOfBibitem
\bibitem[van~de Krol and Liang(2013)]{van_de_krol_n-si/n-fe2o3_2013}
van~de Krol,~R.; Liang,~Y. \emph{{CHIMIA} International Journal for Chemistry}
  \textbf{2013}, \emph{67}, 168--171\relax
\mciteBstWouldAddEndPuncttrue
\mciteSetBstMidEndSepPunct{\mcitedefaultmidpunct}
{\mcitedefaultendpunct}{\mcitedefaultseppunct}\relax
\EndOfBibitem
\bibitem[Sivula(2013)]{sivula_solar--chemical_2013}
Sivula,~K. \emph{{CHIMIA} International Journal for Chemistry} \textbf{2013},
  \emph{67}, 155--161\relax
\mciteBstWouldAddEndPuncttrue
\mciteSetBstMidEndSepPunct{\mcitedefaultmidpunct}
{\mcitedefaultendpunct}{\mcitedefaultseppunct}\relax
\EndOfBibitem
\bibitem[Abdi et~al.(2013)Abdi, Han, Smets, Zeman, Dam, and van~de
  Krol]{abdi_efficient_2013-1}
Abdi,~F.~F.; Han,~L.; Smets,~A. H.~M.; Zeman,~M.; Dam,~B.; van~de Krol,~R.
  \emph{Nature Communications} \textbf{2013}, \emph{4}, year\relax
\mciteBstWouldAddEndPuncttrue
\mciteSetBstMidEndSepPunct{\mcitedefaultmidpunct}
{\mcitedefaultendpunct}{\mcitedefaultseppunct}\relax
\EndOfBibitem
\bibitem[Krol and Gr{\"a}tzel(2011)]{krol_photoelectrochemical_2011}
Krol,~R. V.~D.; Gr{\"a}tzel,~M. \emph{Photoelectrochemical Hydrogen
  Production};
\newblock Springer, 2011\relax
\mciteBstWouldAddEndPuncttrue
\mciteSetBstMidEndSepPunct{\mcitedefaultmidpunct}
{\mcitedefaultendpunct}{\mcitedefaultseppunct}\relax
\EndOfBibitem
\bibitem[G{\"a}rtner(1959)]{gartner_depletion-layer_1959}
G{\"a}rtner,~W.~W. \emph{Physical Review} \textbf{1959}, \emph{116},
  84--87\relax
\mciteBstWouldAddEndPuncttrue
\mciteSetBstMidEndSepPunct{\mcitedefaultmidpunct}
{\mcitedefaultendpunct}{\mcitedefaultseppunct}\relax
\EndOfBibitem
\bibitem[Wilson(1977)]{wilson_model_1977}
Wilson,~R.~H. \emph{Journal of Applied Physics} \textbf{1977}, \emph{48},
  4292--4297\relax
\mciteBstWouldAddEndPuncttrue
\mciteSetBstMidEndSepPunct{\mcitedefaultmidpunct}
{\mcitedefaultendpunct}{\mcitedefaultseppunct}\relax
\EndOfBibitem
\bibitem[Reichman(1980)]{reichman_currentvoltage_1980}
Reichman,~J. \emph{Applied Physics Letters} \textbf{1980}, \emph{36},
  574--577\relax
\mciteBstWouldAddEndPuncttrue
\mciteSetBstMidEndSepPunct{\mcitedefaultmidpunct}
{\mcitedefaultendpunct}{\mcitedefaultseppunct}\relax
\EndOfBibitem
\bibitem[Andrade et~al.(2011)Andrade, Lopes, Ribeiro, and
  Mendes]{andrade_transient_2011}
Andrade,~L.; Lopes,~T.; Ribeiro,~H.~A.; Mendes,~A. \emph{International Journal
  of Hydrogen Energy} \textbf{2011}, \emph{36}, 175--188\relax
\mciteBstWouldAddEndPuncttrue
\mciteSetBstMidEndSepPunct{\mcitedefaultmidpunct}
{\mcitedefaultendpunct}{\mcitedefaultseppunct}\relax
\EndOfBibitem
\bibitem[Foley et~al.(2012)Foley, Price, Feldblyum, and
  Maldonado]{foley_analysis_2012}
Foley,~J.~M.; Price,~M.~J.; Feldblyum,~J.~I.; Maldonado,~S. \emph{Energy \&
  Environmental Science} \textbf{2012}, \emph{5}, 5203--5220\relax
\mciteBstWouldAddEndPuncttrue
\mciteSetBstMidEndSepPunct{\mcitedefaultmidpunct}
{\mcitedefaultendpunct}{\mcitedefaultseppunct}\relax
\EndOfBibitem
\bibitem[Peter et~al.(1984)Peter, Li, and Peat]{peter_surface_1984}
Peter,~L.; Li,~J.; Peat,~R. \emph{Journal of Electroanalytical Chemistry and
  Interfacial Electrochemistry} \textbf{1984}, \emph{165}, 29--40\relax
\mciteBstWouldAddEndPuncttrue
\mciteSetBstMidEndSepPunct{\mcitedefaultmidpunct}
{\mcitedefaultendpunct}{\mcitedefaultseppunct}\relax
\EndOfBibitem
\bibitem[Klahr et~al.(2012)Klahr, Gimenez, Fabregat-Santiago, Hamann, and
  Bisquert]{klahr_water_2012}
Klahr,~B.; Gimenez,~S.; Fabregat-Santiago,~F.; Hamann,~T.; Bisquert,~J.
  \emph{J. Am. Chem. Soc.} \textbf{2012}, \emph{134}, 4294--4302\relax
\mciteBstWouldAddEndPuncttrue
\mciteSetBstMidEndSepPunct{\mcitedefaultmidpunct}
{\mcitedefaultendpunct}{\mcitedefaultseppunct}\relax
\EndOfBibitem
\bibitem[Bertoluzzi and Bisquert(2012)]{bertoluzzi_equivalent_2012}
Bertoluzzi,~L.; Bisquert,~J. \emph{The Journal of Physical Chemistry Letters}
  \textbf{2012},  2517--2522\relax
\mciteBstWouldAddEndPuncttrue
\mciteSetBstMidEndSepPunct{\mcitedefaultmidpunct}
{\mcitedefaultendpunct}{\mcitedefaultseppunct}\relax
\EndOfBibitem
\bibitem[Carver et~al.(2012)Carver, Ulissi, Ong, Dennison, Kelsall, and
  Hellgardt]{carver_modelling_2012}
Carver,~C.; Ulissi,~Z.; Ong,~C.; Dennison,~S.; Kelsall,~G.; Hellgardt,~K.
  \emph{International Journal of Hydrogen Energy} \textbf{2012}, \emph{37},
  2911--2923\relax
\mciteBstWouldAddEndPuncttrue
\mciteSetBstMidEndSepPunct{\mcitedefaultmidpunct}
{\mcitedefaultendpunct}{\mcitedefaultseppunct}\relax
\EndOfBibitem
\bibitem[Haussener et~al.(2013)Haussener, Hu, Xiang, Weber, and
  Lewis]{haussener_simulations_2013}
Haussener,~S.; Hu,~S.; Xiang,~C.; Weber,~A.~Z.; Lewis,~N. \emph{Energy \&
  Environmental Science} \textbf{2013}\relax
\mciteBstWouldAddEndPuncttrue
\mciteSetBstMidEndSepPunct{\mcitedefaultmidpunct}
{\mcitedefaultendpunct}{\mcitedefaultseppunct}\relax
\EndOfBibitem
\bibitem[Salvador(2001)]{salvador_semiconductors_2001}
Salvador,~P. \emph{The Journal of Physical Chemistry B} \textbf{2001},
  \emph{105}, 6128--6141\relax
\mciteBstWouldAddEndPuncttrue
\mciteSetBstMidEndSepPunct{\mcitedefaultmidpunct}
{\mcitedefaultendpunct}{\mcitedefaultseppunct}\relax
\EndOfBibitem
\bibitem[Memming(2008)]{memming_semiconductor_2008}
Memming,~R. \emph{Semiconductor Electrochemistry};
\newblock John Wiley \& Sons, 2008\relax
\mciteBstWouldAddEndPuncttrue
\mciteSetBstMidEndSepPunct{\mcitedefaultmidpunct}
{\mcitedefaultendpunct}{\mcitedefaultseppunct}\relax
\EndOfBibitem
\bibitem[Peter(2013)]{peter_energetics_2013}
Peter,~L.~M. \emph{Journal of Solid State Electrochemistry} \textbf{2013},
  \emph{17}, 315--326\relax
\mciteBstWouldAddEndPuncttrue
\mciteSetBstMidEndSepPunct{\mcitedefaultmidpunct}
{\mcitedefaultendpunct}{\mcitedefaultseppunct}\relax
\EndOfBibitem
\bibitem[Cendula()]{cendula_model_????}
Cendula,~P. \emph{The model is available freely on the internet}.
  \url{http://icp.zhaw.ch/PEC}\relax
\mciteBstWouldAddEndPuncttrue
\mciteSetBstMidEndSepPunct{\mcitedefaultmidpunct}
{\mcitedefaultendpunct}{\mcitedefaultseppunct}\relax
\EndOfBibitem
\bibitem[Dotan et~al.(2011)Dotan, Sivula, Gr{\"a}tzel, Rothschild, and
  Warren]{dotan_probing_2011}
Dotan,~H.; Sivula,~K.; Gr{\"a}tzel,~M.; Rothschild,~A.; Warren,~S.~C.
  \emph{Energy \& Environmental Science} \textbf{2011}, \emph{4}, 958\relax
\mciteBstWouldAddEndPuncttrue
\mciteSetBstMidEndSepPunct{\mcitedefaultmidpunct}
{\mcitedefaultendpunct}{\mcitedefaultseppunct}\relax
\EndOfBibitem
\bibitem[Klahr and Hamann(2011)]{klahr_voltage_2011}
Klahr,~B.~M.; Hamann,~T.~W. \emph{Applied Physics Letters} \textbf{2011},
  \emph{99}, 063508--063508--3\relax
\mciteBstWouldAddEndPuncttrue
\mciteSetBstMidEndSepPunct{\mcitedefaultmidpunct}
{\mcitedefaultendpunct}{\mcitedefaultseppunct}\relax
\EndOfBibitem
\bibitem[Bisquert et~al.(2013)Bisquert, Cendula, Bertoluzzi, and
  Gimenez]{bisquert_energy_2013}
Bisquert,~J.; Cendula,~P.; Bertoluzzi,~L.; Gimenez,~S. \emph{The Journal of
  Physical Chemistry Letters} \textbf{2013},  205--207\relax
\mciteBstWouldAddEndPuncttrue
\mciteSetBstMidEndSepPunct{\mcitedefaultmidpunct}
{\mcitedefaultendpunct}{\mcitedefaultseppunct}\relax
\EndOfBibitem
\bibitem[{NREL}(2012)]{nrel_solar_2012}
{NREL}, \emph{Solar Spectral Irradiance: Air Mass 1.5, downloaded March 2012},
  2012. \url{http://rredc.nrel.gov/solar/spectra/am1.5/}\relax
\mciteBstWouldAddEndPuncttrue
\mciteSetBstMidEndSepPunct{\mcitedefaultmidpunct}
{\mcitedefaultendpunct}{\mcitedefaultseppunct}\relax
\EndOfBibitem
\bibitem[Tan et~al.(1994)Tan, Laibinis, Nguyen, Kesselman, Stanton, and
  Lewis]{tan_principles_1994}
Tan,~M.~X.; Laibinis,~P.~E.; Nguyen,~S.~T.; Kesselman,~J.~M.; Stanton,~C.~E.;
  Lewis,~N.~S. Principles and Applications of Semiconductor
  Photoelectrochemistry. In \emph{Progress in Inorganic Chemistry};
  Karlin,~K.~D., Ed.;
\newblock John Wiley \& Sons, Inc., 1994;
\newblock pp 21--144\relax
\mciteBstWouldAddEndPuncttrue
\mciteSetBstMidEndSepPunct{\mcitedefaultmidpunct}
{\mcitedefaultendpunct}{\mcitedefaultseppunct}\relax
\EndOfBibitem
\bibitem[Markowich et~al.(1990)Markowich, Ringhofer, and
  Schmeiser]{markowich_semiconductor_1990}
Markowich,~P.~A.; Ringhofer,~C.~A.; Schmeiser,~C. \emph{Semiconductor
  equations};
\newblock Springer-Verlag New York, Inc.: New York, {NY}, {USA}, 1990\relax
\mciteBstWouldAddEndPuncttrue
\mciteSetBstMidEndSepPunct{\mcitedefaultmidpunct}
{\mcitedefaultendpunct}{\mcitedefaultseppunct}\relax
\EndOfBibitem
\bibitem[Kay et~al.(2006)Kay, Cesar, and Gr{\"a}tzel]{kay_new_2006}
Kay,~A.; Cesar,~I.; Gr{\"a}tzel,~M. \emph{J. Am. Chem. Soc.} \textbf{2006},
  \emph{128}, 15714--15721\relax
\mciteBstWouldAddEndPuncttrue
\mciteSetBstMidEndSepPunct{\mcitedefaultmidpunct}
{\mcitedefaultendpunct}{\mcitedefaultseppunct}\relax
\EndOfBibitem
\bibitem[Chen et~al.(2013)Chen, Deutsch, Dinh, Domen, Emery, Forman, Gaillard,
  Garland, Heske, Jaramillo, Kleiman-Shwarsctein, Miller, Takanabe, and
  Turner]{chen_efficiency_2013}
Chen,~Z.; Deutsch,~T.~G.; Dinh,~H.~N.; Domen,~K.; Emery,~K.; Forman,~A.~J.;
  Gaillard,~N.; Garland,~R.; Heske,~C.; Jaramillo,~T.~F.;
  Kleiman-Shwarsctein,~A.; Miller,~E.; Takanabe,~K.; Turner,~J. Efficiency
  Definitions in the Field of {PEC}. In \emph{Photoelectrochemical Water
  Splitting};
\newblock {SpringerBriefs} in Energy;
\newblock Springer New York, 2013;
\newblock pp 7--16\relax
\mciteBstWouldAddEndPuncttrue
\mciteSetBstMidEndSepPunct{\mcitedefaultmidpunct}
{\mcitedefaultendpunct}{\mcitedefaultseppunct}\relax
\EndOfBibitem
\bibitem[Nozik(1978)]{nozik_photoelectrochemistry:_1978}
Nozik,~A.~J. \emph{Annual Review of Physical Chemistry} \textbf{1978},
  \emph{29}, 189--222\relax
\mciteBstWouldAddEndPuncttrue
\mciteSetBstMidEndSepPunct{\mcitedefaultmidpunct}
{\mcitedefaultendpunct}{\mcitedefaultseppunct}\relax
\EndOfBibitem
\bibitem[Gr{\"a}tzel(2001)]{gratzel_photoelectrochemical_2001}
Gr{\"a}tzel,~M. \emph{Nature} \textbf{2001}, \emph{414}, 338--344\relax
\mciteBstWouldAddEndPuncttrue
\mciteSetBstMidEndSepPunct{\mcitedefaultmidpunct}
{\mcitedefaultendpunct}{\mcitedefaultseppunct}\relax
\EndOfBibitem
\bibitem[Krol et~al.(2008)Krol, Liang, and Schoonman]{krol_solar_2008}
Krol,~R. v.~d.; Liang,~Y.; Schoonman,~J. \emph{J. Mater. Chem.} \textbf{2008},
  \emph{18}, 2311--2320\relax
\mciteBstWouldAddEndPuncttrue
\mciteSetBstMidEndSepPunct{\mcitedefaultmidpunct}
{\mcitedefaultendpunct}{\mcitedefaultseppunct}\relax
\EndOfBibitem
\bibitem[Hodes(2012)]{hodes_photoelectrochemical_2012}
Hodes,~G. \emph{The Journal of Physical Chemistry Letters} \textbf{2012},
  \emph{3}, 1208--1213\relax
\mciteBstWouldAddEndPuncttrue
\mciteSetBstMidEndSepPunct{\mcitedefaultmidpunct}
{\mcitedefaultendpunct}{\mcitedefaultseppunct}\relax
\EndOfBibitem
\bibitem[Morrison(1980)]{morrison_electrochemistry_1980}
Morrison,~S.~R. \emph{Electrochemistry at semiconductor and oxidized metal
  electrodes};
\newblock Plenum Press, 1980\relax
\mciteBstWouldAddEndPuncttrue
\mciteSetBstMidEndSepPunct{\mcitedefaultmidpunct}
{\mcitedefaultendpunct}{\mcitedefaultseppunct}\relax
\EndOfBibitem
\bibitem[Tilley et~al.(2010)Tilley, Cornuz, Sivula, and
  Gr{\"a}tzel]{tilley_lightinduced_2010}
Tilley,~S.~D.; Cornuz,~M.; Sivula,~K.; Gr{\"a}tzel,~M. \emph{Angewandte Chemie}
  \textbf{2010}, \emph{122}, 6549--6552\relax
\mciteBstWouldAddEndPuncttrue
\mciteSetBstMidEndSepPunct{\mcitedefaultmidpunct}
{\mcitedefaultendpunct}{\mcitedefaultseppunct}\relax
\EndOfBibitem
\bibitem[Klahr et~al.(2012)Klahr, Gimenez, Fabregat-Santiago, Bisquert, and
  Hamann]{klahr_electrochemical_2012}
Klahr,~B.; Gimenez,~S.; Fabregat-Santiago,~F.; Bisquert,~J.; Hamann,~T.~W.
  \emph{Energy \& Environmental Science} \textbf{2012}, \emph{5},
  7626--7636\relax
\mciteBstWouldAddEndPuncttrue
\mciteSetBstMidEndSepPunct{\mcitedefaultmidpunct}
{\mcitedefaultendpunct}{\mcitedefaultseppunct}\relax
\EndOfBibitem
\bibitem[Paracchino et~al.(2011)Paracchino, Laporte, Sivula, Gr{\"a}tzel, and
  Thimsen]{paracchino_highly_2011}
Paracchino,~A.; Laporte,~V.; Sivula,~K.; Gr{\"a}tzel,~M.; Thimsen,~E. \emph{Nat
  Mater} \textbf{2011}, \emph{10}, 456--461\relax
\mciteBstWouldAddEndPuncttrue
\mciteSetBstMidEndSepPunct{\mcitedefaultmidpunct}
{\mcitedefaultendpunct}{\mcitedefaultseppunct}\relax
\EndOfBibitem
\bibitem[Paracchino et~al.(2012)Paracchino, Mathews, Hisatomi, Stefik, Tilley,
  and Gr{\"a}tzel]{paracchino_ultrathin_2012}
Paracchino,~A.; Mathews,~N.; Hisatomi,~T.; Stefik,~M.; Tilley,~S.~D.;
  Gr{\"a}tzel,~M. \emph{Energy \& Environmental Science} \textbf{2012},
  \emph{5}, 8673\relax
\mciteBstWouldAddEndPuncttrue
\mciteSetBstMidEndSepPunct{\mcitedefaultmidpunct}
{\mcitedefaultendpunct}{\mcitedefaultseppunct}\relax
\EndOfBibitem
\bibitem[Tilley et~al.(2013)Tilley, Schreier, Azevedo, Stefik, and
  Graetzel]{tilley_ruthenium_2013}
Tilley,~S.~D.; Schreier,~M.; Azevedo,~J.; Stefik,~M.; Graetzel,~M.
  \emph{Advanced Functional Materials} \textbf{2013},
  n/a{\textendash}n/a\relax
\mciteBstWouldAddEndPuncttrue
\mciteSetBstMidEndSepPunct{\mcitedefaultmidpunct}
{\mcitedefaultendpunct}{\mcitedefaultseppunct}\relax
\EndOfBibitem
\bibitem[Sivula(2013)]{sivula_metal_2013}
Sivula,~K. \emph{The Journal of Physical Chemistry Letters} \textbf{2013},
  \emph{4}, 1624--1633\relax
\mciteBstWouldAddEndPuncttrue
\mciteSetBstMidEndSepPunct{\mcitedefaultmidpunct}
{\mcitedefaultendpunct}{\mcitedefaultseppunct}\relax
\EndOfBibitem
\bibitem[Upul~Wijayantha et~al.(2011)Upul~Wijayantha, Saremi-Yarahmadi, and
  Peter]{upul_wijayantha_kinetics_2011}
Upul~Wijayantha,~K.; Saremi-Yarahmadi,~S.; Peter,~L.~M. \emph{Physical
  Chemistry Chemical Physics} \textbf{2011}, \emph{13}, 5264\relax
\mciteBstWouldAddEndPuncttrue
\mciteSetBstMidEndSepPunct{\mcitedefaultmidpunct}
{\mcitedefaultendpunct}{\mcitedefaultseppunct}\relax
\EndOfBibitem
\bibitem[Paracchino et~al.(2012)Paracchino, Brauer, Moser, Thimsen, and
  Graetzel]{paracchino_synthesis_2012}
Paracchino,~A.; Brauer,~J.~C.; Moser,~J.-E.; Thimsen,~E.; Graetzel,~M.
  \emph{The Journal of Physical Chemistry C} \textbf{2012}, \emph{116},
  7341--7350\relax
\mciteBstWouldAddEndPuncttrue
\mciteSetBstMidEndSepPunct{\mcitedefaultmidpunct}
{\mcitedefaultendpunct}{\mcitedefaultseppunct}\relax
\EndOfBibitem
\bibitem[Xu and Schoonen(2000)]{xu_absolute_2000}
Xu,~Y.; Schoonen,~M. A.~A. \emph{American Mineralogist} \textbf{2000},
  \emph{85}, 543--556\relax
\mciteBstWouldAddEndPuncttrue
\mciteSetBstMidEndSepPunct{\mcitedefaultmidpunct}
{\mcitedefaultendpunct}{\mcitedefaultseppunct}\relax
\EndOfBibitem
\bibitem[Niu et~al.(2010)Niu, Huang, Cui, Huang, Yu, and
  Wang]{niu_hydrothermal_2010}
Niu,~M.; Huang,~F.; Cui,~L.; Huang,~P.; Yu,~Y.; Wang,~Y. \emph{{ACS} Nano}
  \textbf{2010}, \emph{4}, 681--688\relax
\mciteBstWouldAddEndPuncttrue
\mciteSetBstMidEndSepPunct{\mcitedefaultmidpunct}
{\mcitedefaultendpunct}{\mcitedefaultseppunct}\relax
\EndOfBibitem
\bibitem[Morin(1954)]{morin_electrical_1954}
Morin,~F.~J. \emph{Physical Review} \textbf{1954}, \emph{93}, 1195--1199\relax
\mciteBstWouldAddEndPuncttrue
\mciteSetBstMidEndSepPunct{\mcitedefaultmidpunct}
{\mcitedefaultendpunct}{\mcitedefaultseppunct}\relax
\EndOfBibitem
\bibitem[Cesar et~al.(2008)Cesar, Sivula, Kay, Zboril, and
  Gr{\"a}tzel]{cesar_influence_2008}
Cesar,~I.; Sivula,~K.; Kay,~A.; Zboril,~R.; Gr{\"a}tzel,~M. \emph{J. Phys.
  Chem. C} \textbf{2008}, \emph{113}, 772--782\relax
\mciteBstWouldAddEndPuncttrue
\mciteSetBstMidEndSepPunct{\mcitedefaultmidpunct}
{\mcitedefaultendpunct}{\mcitedefaultseppunct}\relax
\EndOfBibitem
\bibitem[Bosman and van Daal(1970)]{bosman_small-polaron_1970}
Bosman,~A.; van Daal,~H. \emph{Advances in Physics} \textbf{1970}, \emph{19},
  1--117\relax
\mciteBstWouldAddEndPuncttrue
\mciteSetBstMidEndSepPunct{\mcitedefaultmidpunct}
{\mcitedefaultendpunct}{\mcitedefaultseppunct}\relax
\EndOfBibitem
\end{mcitethebibliography}

\end{document}